\documentclass[aps,showpacs,superscriptaddress,preprint]{revtex4-2}
\usepackage{latexsym,graphicx,epsfig,psfrag}

\usepackage{epstopdf}
\usepackage{amssymb,amsmath,amsxtra,amsfonts}
\usepackage{bm}

\usepackage{longtable}
\usepackage{multirow}
\usepackage{booktabs}
\usepackage{array}
\usepackage{wrapfig}
\usepackage{color}
\usepackage{titlesec}
\usepackage[title]{appendix}

\titleformat{\section}{\large\bfseries}{\thesection}{1em}{}

\newcommand{\bea}{\begin{eqnarray}}
\newcommand{\ena}{\end{eqnarray}}
\newcommand{\be}{\begin{equation}}
\newcommand{\en}{\end{equation}}
\newcommand{\nn}{\nonumber\\}
\newcommand{\ed}{\end{document}} 
\newcommand{\Tr}{\mbox{\rm{tr}}}

\begin{document}

\title{Radiative decays \bm{$D^*_{(s)}\to D_{(s)}\gamma$} in covariant confined quark model} 

\author{C.~T.~Tran}
\email{thangtc@hcmute.edu.vn (corresponding author)}
\affiliation{
	\hbox{Department of Physics, HCMC University of Technology and Education, }\\
	Vo Van Ngan 1, 700000 Ho Chi Minh City, Vietnam}

\author{M. A. Ivanov}
\email{ivanovm@theor.jinr.ru}
\affiliation{Bogoliubov Laboratory of Theoretical Physics, 
Joint Institute for Nuclear Research, 141980 Dubna, Russia}

\author{P.~Santorelli}
\email{Pietro.Santorelli@na.infn.it}
\affiliation{
	Dipartimento di Fisica ``E.~Pancini'', Universit\`a di Napoli Federico II,
	\hbox{ Complesso Universitario di Monte S.~Angelo,
		Via Cintia, Edificio 6, 80126 Napoli, Italy}}
\affiliation{
	\hbox{Istituto Nazionale di Fisica Nucleare, 
		Sezione di Napoli, 80126 Napoli, Italy}}

\author{Q.~C.~Vo}
\affiliation{
	\hbox{University of Science, 700000 Ho Chi Minh City, Vietnam}}

\begin{abstract}
Radiative decays $D^*_{(s)}\to D_{(s)}\gamma$ are revisited in light of new experimental data from the \textit{BABAR} and BESIII collaborations. The radiative couplings $g_{D^*D\gamma}$ encoding nonperturbative QCD effects are calculated in the framework of the covariant confined quark model developed by us. We compare our results with other theoretical studies and experimental data. The couplings (in $\textrm{GeV}^{-1}$)  $|g_{D^{*+}D^+\gamma}| = 0.45(9)$ and $|g_{D^{*0}D^0\gamma}| = 1.72(34)$ calculated in our model agree with the corresponding experimental data $|g_{D^{*+}D^+\gamma}|=0.47(7)$ and $|g_{D^{*0}D^0\gamma}|=1.77(16)$. The most interesting case is the decay $D^*_s\to D_s\gamma$,  
for which a recent prediction based on light-cone sum rules at next-to-leading order $|g_{D^*_s D_s\gamma}|=0.60(19)$ deviates from the first (and only to date) lattice QCD result $|g_{D^*_s D_s\gamma}|=0.11(2)$ at nearly  $3\sigma$.
Our calculation yields $|g_{D^*_s D_s\gamma}|=0.29(6)$,
which falls somehow between the two mentioned results, although it is larger than those predicted in other studies using quark models or QCD sum rules.  
\end{abstract}


\maketitle
\newpage

\section{Introduction}
\label{sec:intro}
Decays of $D^*$ mesons are dominated by the radiative  $D^*\to D\gamma$ ($\gamma$ emission) and strong $D^*\to D\pi$ ($\pi$ emission) processes. The hadronic effects in these decays are encoded in the magnetic moment coupling $g_{D^*D\gamma}$ and axial coupling $g_{D^*D\pi}$, respectively.  Owing to experimental difficulties, data concerning their width are rare. The first measurement of the width of a vector $D$ meson was achieved by the CLEO collaboration for the charged $D^{*+}$ meson. They observed $\Gamma(D^{*+})=(96\pm 4\pm 22)\,\textrm{keV}$~\cite{CLEO:2001sxb}, whose uncertainties are statistical and systematic. In 2013, the \textit{BABAR} collaboration obtained the width with much higher precision: $\Gamma(D^{*+})=(83.3\pm 1.2\pm 1.4)\,\textrm{keV}$~\cite{BaBar:2013thi}. In 2014, the BESIII collaboration reported the measurements of the branching fractions of the $D^{*0}$ decays to be $\mathcal{B}(D^{*0}\to D^0\pi^0)=(65.5\pm 0.8\pm 0.5)\%$ and $\mathcal{B}(D^{*0}\to D^0\gamma)=(34.5\pm 0.8\pm 0.5)\%$, assuming the dominance of these two channels~\cite{BESIII:2014rqs}. These are the most precise measurements of the branching fractions to date. Recently, the BESIII collaboration has announced the measurement of the branching of $D_s^{*+}\to D_s^+\pi^0$ relative to $D_s^{*+}\to D_s^+\gamma$ to be $(6.16\pm 0.43\pm 0.19)\%$~\cite{BESIII:2022kbd}. This measurement confirmed the previous BABAR's result and made it more accurate~\cite{BaBar:2005wmf}. It should be noted that the decay $D_s^{*+}\to D_s^+\pi^0$ violates isospin symmetry and is highly suppressed with respect to the dominating mode $D_s^{*+}\to D_s^+\gamma$. The mechanism of $D_s^{*+}\to D_s^+\pi^0$ is still an open question. However, it is evidently not the same $\pi$ emission as for $D^{*+}\to (D\pi)^+$ and $D^{*0}\to D^0\pi^0$. It can be concluded that, in light of these new experimental data, as well as data to be reported in the near future from BESIII, the study of $D^*$-meson radiative decays is fruitful and promising.     

Radiative decays of $D^*$ meson can shed light on the dynamics of many hadronic processes. They allow testing various non-perturbative approaches in QCD. This explains the large amount of theoretical attempts to calculate these decays, especially prior to the CLEO's first measurement of $\Gamma(D^*)$. Radiative decays $D^*\to D\gamma$ have been studied in the framework of QCD sum rules (QCDSR)~\cite{Eletsky:1984qs, Aliev:1994nq, Dosch:1995kw, Zhu:1996qy}, light cone sum rules (LCSR)~\cite{Aliev:1995zlh,Li:2020rcg}, and various quark models~\cite{ODonnell:1994qrt, Ivanov:1994ji, Deandrea:1998uz, Colangelo:1994jc, Deng:2013uca, Orsland:1998de, Jaus:1996np,  Goity:2000dk, Ebert:2002xz, Choi:2007se, Cheung:2014cka}. Lattice QCD (LQCD) calculation has been carried out for decays $D^*\to D\gamma$~\cite{Becirevic:2009xp} and $D^*_s\to D_s\gamma$~\cite{Donald:2013sra}. However, the theoretical uncertainties in these calculations are still significant. Recently, the first complete next-to-leading order (NLO) computation of the decays $D^*\to D\gamma$ using LCSR has been reported~\cite{Pullin:2021ebn}. In Ref.~\cite{Pullin:2021ebn}, the authors pointed out some differences between the first LQCD prediction~\cite{Donald:2013sra} of $g_{D_s^*D_s\gamma}=0.11(2)\,\textrm{GeV}^{-1}$ and the current experimental value $g_{D^*D\gamma}=-0.47(7)\,\textrm{GeV}^{-1}$, assuming the level of $D$-spin breaking to be no more than 20--30\%. Moreover, the LQCD result~\cite{Donald:2013sra} predicted the decay width to be $\Gamma(D_s^*\to D_s\gamma)=0.066(26)$~keV.
		This value disagrees with the result obtained from the light front quark model $\Gamma(D_s^*\to D_s\gamma)=0.18(1)$~keV~\cite{Choi:2007se} by $4\sigma$, and with the result from QCDSR $\Gamma(D_s^*\to D_s\gamma)=0.25(8)$~keV~\cite{Aliev:1994nq} by $2\sigma$. The central value of this prediction 
is one order of magnitude smaller than most of other predictions available in the literature.  Given that $D^*_s\to D_s\gamma$ dominates the decay of the $D^*_s$ meson, it is interesting to examine this case more carefully. 

In this paper, we revisit the radiative decays of the $D^*$ mesons in the framework of the covariant confined quark model (CCQM) developed previously by our group. This is a model based on quantum field theory in which the interaction between the constituent quarks and corresponding hadronic bound state is described by a Lagrangian in the form of a hadron field coupled to a quark current. Our model has the advantages of treating multiquark states in a similar manner and obtaining observables in the full physical range without any extrapolation. We aimed to provide predictions for both $D^*$ and $D_s^*$ decays; nevertheless, we further focused on the decay $D^*_s\to D_s\gamma$, for which theoretical predictions still vary significantly and experimental data on its width are expected to be available soon. 

The rest of the paper is organized as follows. In Sec.~\ref{sec:CCQM}, we briefly introduce the covariant confined quark model. The calculation of the radiative decays $D^*_{(s)}\to D\gamma$ using our model is demonstrated in Sec.~\ref{sec:decay}. Numerical results are presented in Sec.~\ref{sec:result}, where we also provide a detailed comparison of our results with experimental data and other theoretical studies. Finally, a brief summary is given in Sec.~\ref{sec:sum}.


\section{Covariant Confined Quark Model}
\label{sec:CCQM}
The covariant confined quark model (CCQM) is a model based on quantum field theory. In this model, hadronic bound states are described by quantum fields that interact with their constituent quarks via an interaction Lagrangian. In the case of a meson $M(q_1\bar{q}_2)$, this Lagrangian has the form
\begin{equation}
	\label{eq:LInt}
	\mathcal{L}_{\mathrm{int}}(x) =g_M M(x)J(x)+\mathrm{H.c.},\qquad
	J(x) = \int dx_1\int dx_2 F_M(x;x_1,x_2)[\bar{q}_2(x_2)\Gamma_M q_1(x_1)],
\end{equation}  
where $\Gamma_M$ is the relevant Dirac matrix and $g_M$ is the meson-quark coupling constant. The vertex function $F_M(x;x_1,x_2)$ effectively represents the finite size of the meson and relative quark-hadron position. The vertex function has to obey the identity $F_M(x+a;x_1+a,x_2+a)=F_M(x;x_1,x_2)$ for any given four-vector $a$ so as to satisfy the translational invariance.
Its form is chosen as
\begin{equation}
	F_M(x;x_1,x_2) = \delta^{(4)}(x-\omega_1 x_1-\omega_2 x_2)\Phi_M[(x_1-x_2)^2],
\end{equation} 
where $\omega_{i}=m_{q_i}/(m_{q_1}+m_{q_2})$ with $m_{q_i}$ being the constituent quark mass; therefore, $\omega_1+\omega_2=1$. This chosen form satisfies the translational invariance mentioned above. Besides, it corresponds to an effective description of a meson. The Dirac delta function effectively represents the barycenter of the two-quark system. The function $\Phi_M[(x_1-x_2)^2]$ depends on the distance between the two quarks, which can be understood as the effective size of the meson. The Fourier transform of the function $\Phi_M[(x_1-x_2)^2]$, which we denote as $\widetilde{\Phi}_M(-k^2)$, can be calculated, in principle, from the solutions of the Bethe-Salpeter equation for meson bound states. In a series of previous papers of us~\cite{Anikin:1995cf, Ivanov:1997ug, Faessler:2003yf}, after trying various forms for the vertex function, we found that the hadronic observables are insensitive to the details of the functional form of the quark-hadron vertex function. This observation is used as a guiding principle, and the function $\Phi_M[(x_1-x_2)^2]$ is assumed to be Gaussian for simplicity; it is expressed in terms of the momentum representation as
\begin{equation}
	\widetilde{\Phi}_M(-k^2)=\exp(k^2/\Lambda^2_M).
	\label{eq:vertexf}
\end{equation}
The size parameter $\Lambda_M$ of the meson $M$ is one of the free parameters of the model. The minus sign in the argument of the function $\widetilde{\Phi}_M(-k^2)$ emphasizes that we are working in Minkowski space. Given that $k^2$ turns into $-k^2$ in the Euclidean space, the Gaussian function $\exp(k^2/\Lambda^2_M)$ falls off appropriately in the Euclidean region. We stress again that any choice for $\widetilde{\Phi}_M$ is applicable provided it has the sufficiently fast fall-off behavior in the ultraviolet region of the Euclidean space to render the ultraviolet finite of Feynman diagrams.

The normalization of particle-quark vertices is provided by the compositeness 
condition~\cite{Salam:1962ap,Weinberg:1962hj}
\begin{equation}
Z_M = 1 - \widetilde{\Pi}^\prime_M(m^2_M) = 0,
\label{eq:compositeness}
\end{equation}
where $Z_M$ is the wave function renormalization constant of the meson $M$ 
and $\widetilde{\Pi}'_M$ is the derivative of the meson mass function taken on the mass-shell $p^2=m_M^2$. The square root of the renormalization constant $Z_M^{1/2}$ can be interpreted as the matrix element between the physical state and corresponding bare state. When $Z_M=0$, the physical state does not contain the bare state; therefore, it is described as a bound state. The Lagrangian given in Eq.~(\ref{eq:LInt}) describes the interaction of the constituents (quarks) and corresponding physical state (meson), which is the bound state of the constituents. As a result of the interaction, the physical state is dressed, and its mass and wave function have to be renormalized. 

Note that, even before the rise of QCD, it was well understood that describing composite particles such as hadrons within QFT is not easy. In QFT, free fields are quantized by imposing (anti)commutator relations between creation and annihilation operators, which act on vacuum and help construct the asymptotic in- and out-states. Physical processes are described by the S-matrix elements, which are convolutions of propagators. The Lagrangian that describes free fields and their interactions must be renormalized; or in other words, bare or unrenormalized quantities such as wave functions, coupling constants, and mass must be renormalized so that they represent physical (renormalized) ones. When describing a bound state, e.g., the decays of a hadron in a physical process, physical (renormalized) quantities such as its mass must be used. The relation between the bare field and the dressed one can be expressed using the wave function renormalization constant as $\phi_0 = Z^{1/2}\phi_r$. Note that the bare field $\phi_0$ can be eliminated from the Lagrangian by setting its wave function renormalization constant to be zero, i.e., $Z=0$. Jouvet~\cite{Jouvet:1956ii} suggested that the equation $Z=0$ can be used as a compositeness condition. In particular, he proved the equivalence between a four-fermion theory and a Yukawa-type theory if the renormalization constant of the boson field is set to zero. Then, the renormalized boson mass and Yukawa coupling can be expressed in terms of the Fermi constant via the compositeness condition. A more detailed discussion of the compositeness condition is given in the Appendix. 

\begin{figure}[ht]
	\includegraphics[width=0.7\textwidth]{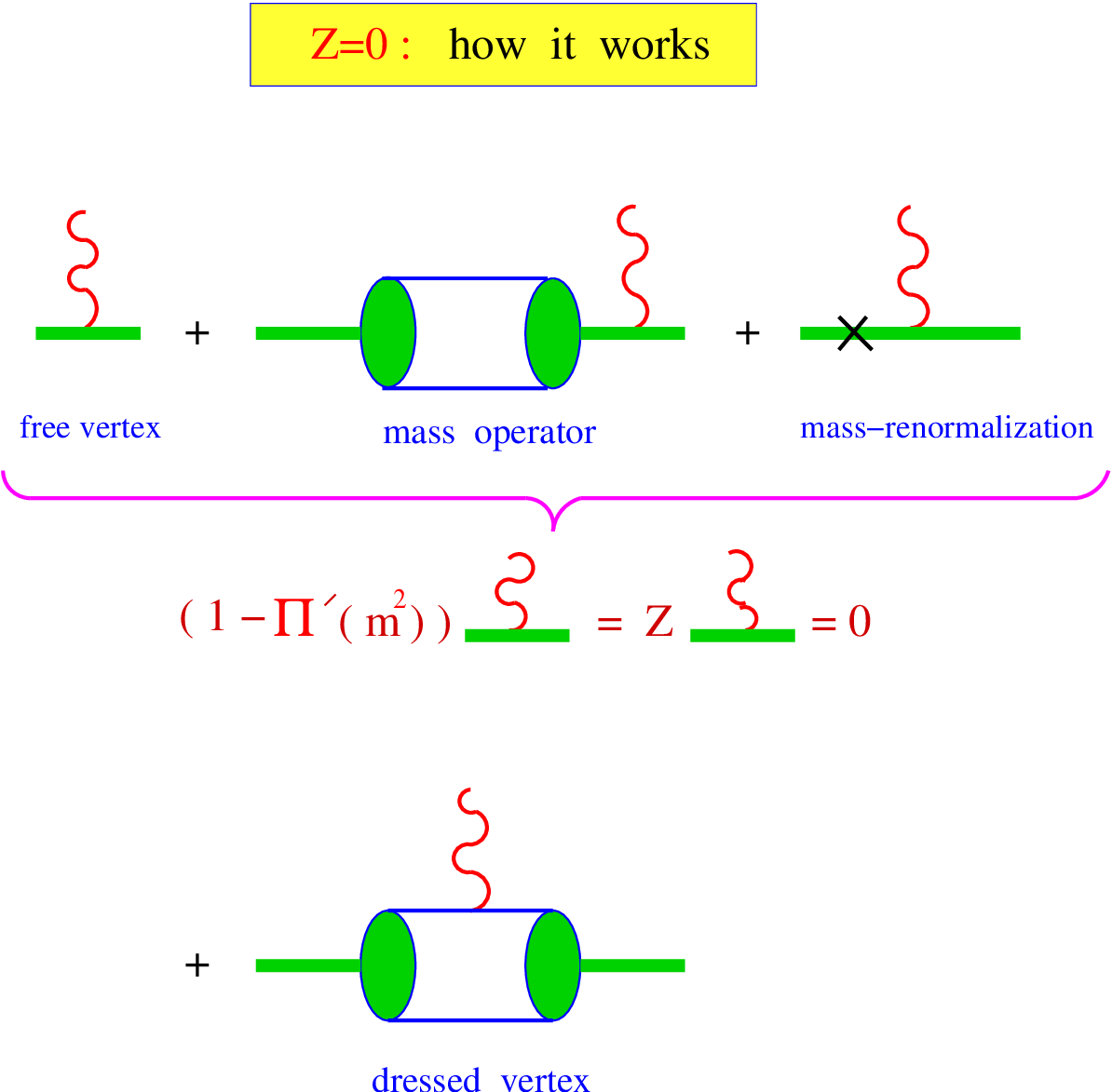}
	\caption{Compositeness condition: how it works.}
	\label{fig:Z=0}
\end{figure}

The compositeness condition effectively excludes the constituent degrees of freedom from the physical state space and avoids double counting. In other words, the constituents exist only in virtual states. 
This point is illustrated in a more intuitive manner in Fig.~\ref{fig:Z=0}.


\begin{figure}[ht]
	\includegraphics[width=0.40\textwidth]{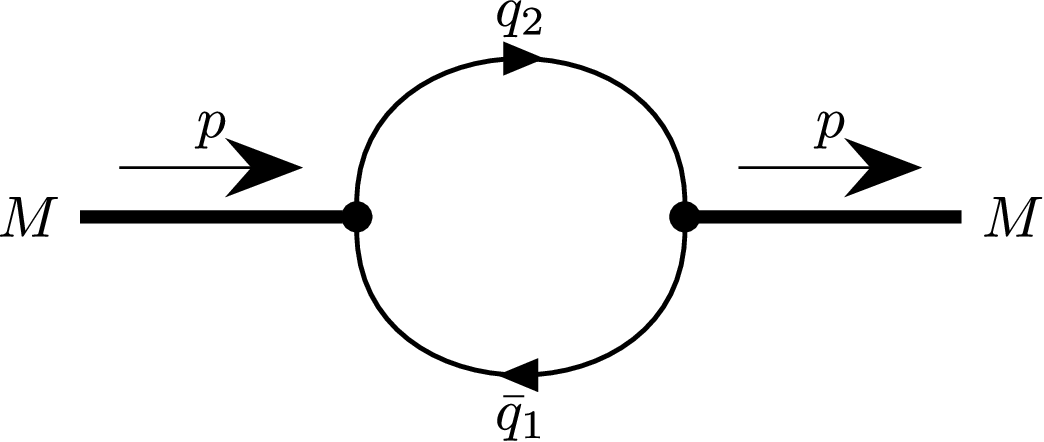}
	\caption{One-loop self-energy diagram for a meson.}
	\label{fig:mass}
\end{figure}

The meson mass function 
in Eq.~(\ref{eq:compositeness}) is defined by the Feynman diagram shown
in Fig.~\ref{fig:mass} and has the following form for a pseudoscalar meson and vector meson, respectively: 
\begin{eqnarray}
\widetilde{\Pi}_P(p) &=& 3g_P^2 \int\!\! \frac{dk}{(2\pi)^4i}\,\widetilde\Phi^2_P \left(-k^2\right)
\Tr\left[ S_1(k+w_1p)\gamma^5 S_2(k-w_2p)\gamma^5 \right],\\
\label{eq:massP}
\widetilde{\Pi}_V(p) &=& g_V^2 \left[g^{\mu\nu} - \frac{p^{\mu}p^{\nu}}{p^2}\right] 
\int\!\! \frac{dk}{(2\pi)^4i}\,\widetilde\Phi^2_V \left(-k^2\right)
\Tr\left[ S_1(k+w_1p)\gamma_{\mu} S_2(k-w_2p)\gamma_{\nu} \right],
\label{eq:massV}
\end{eqnarray}
where we use the free quark propagator
\begin{equation}
S_i(k) = \frac{1}{m_{q_i} - \not\! k - i\epsilon}= \frac{m_{q_i}+\not\! k}{m^2_{q_i} - k^2 - i\epsilon}.
\label{eq:prop}
\end{equation}

It is convenient to calculate the derivatives of the meson mass functions
by using the following identities:
\begin{eqnarray}
	\frac{d}{dp^2} \widetilde\Pi_M(p^2) &=& 
	\frac{1}{2p^2} p^\mu\frac{d}{dp^\mu} \widetilde\Pi_M(p^2),
	\nn
	p^\mu\frac{d}{dp^\mu} S(k + w p)&=& w\, S(k + w p)\not\! p\, S(k + w p).
	\label{eq:identity}
\end{eqnarray}

Then, the  derivatives of the meson mass functions have the form 
\begin{eqnarray}
	\widetilde\Pi'_P(p^2)&=&
	\frac{1}{2p^2}\,\frac{3g^2_P}{4\pi^2}\int\!\! \frac{dk}{4\pi^2i}
	\widetilde\Phi^2_P \left(-k^2\right)
	\nn
	&\times& 
	\Big\{ w_1\,\Tr\left[ S_1(k+w_1p)\not\!p \,S_1(k+w_1p)\gamma^5 
	S_2(k-w_2p)\gamma^5\right]
	\nn
	&& - w_2\, \Tr\left[ S_1(k+w_1p)\gamma^5 S_2(k-w_2p)\not\!p \,
	S_2(k-w_2p)\gamma^5\right]
	\Big\},
	\label{eq:primeP}\\[1.5ex]
	\widetilde\Pi'_V(p^2)&=&
	\frac{1}{2p^2}\,\frac13\left(g^{\mu\nu} - \frac{p^\mu p^\nu}{p^2}\right)
	\frac{3g^2_V}{4\pi^2}
	\int\!\! \frac{dk}{4\pi^2i}\widetilde\Phi^2_V \left(-k^2\right) 
	\nn
	&\times& 
	\Big\{ w_1\,\Tr\left[ S_1(k+w_1p)\not\!p \,S_1(k+w_1p)\gamma_\mu 
	S_2(k-w_2p)\gamma_\nu\right]
	\nn
	&& -w_2\,\Tr\left[ S_1(k+w_1p)\gamma_\mu S_2(k-w_2p)\not\!p \,
	S_2(k-w_2p)\gamma_\nu\right]\Big\}.
	\label{eq:primeV}
\end{eqnarray}

The Fock-Schwinger representation of quark propagators is then used to calculate the loop integrations in Eqs.~(\ref{eq:primeP}) and ~(\ref{eq:primeV}):
\begin{eqnarray}
	S_q (k+w p) &=& \frac{1}{ m_q-\not\! k- w \not\! p } 
	=  \frac{m_q + \not\! k + w \not\! p}{m^2_q - (k+w p)^2}
	\nn
	&=& (m_q + \not\! k + w \not\! p)\int\limits_0^\infty \!\!d\alpha\, 
	e^{-\alpha [m_q^2-(k+w p)^2]}.
	\label{eq:Fock}
\end{eqnarray}
As will be seen later, the Fock-Schwinger representation constitutes an efficient approach for the calculation of tensor loop integrals because loop momenta can be converted into derivatives of the exponent function. 

Loop integrations are performed in the Euclidean space. The transition from the Minkowski space to the Euclidean space is done by the Wick rotation
\begin{equation}
k_0=e^{i\frac{\pi}{2}}k_4=ik_4
\label{eq:Wick}
\end{equation}
so that $k^2=k_0^2-\vec{k}^2=-k_4^2-\vec{k}^2=-k_E^2 \leq 0.$
Simultaneously, all external momenta are rotated, i.e.,
$p_0 \to ip_4$, so that $p^2=-p_E^2 \leq 0$.
Note that the quadratic form in Eq.~(\ref{eq:Fock}) now becomes positive-definite,
\[
m^2_q-(k+w p)^2=m^2_q + (k_E+w p_E)^2>0,
\]
and the integral over $\alpha$ is absolutely convergent.
The Minkowski notation will be kept to avoid 
excessive relabeling; keep in mind that
$k^2 \leq 0$ and $p^2 \leq 0$.

Using the representations of the vertex functions
and quark propagators given by Eqs.~(\ref{eq:vertexf})
and (\ref{eq:Fock}), respectively, we calculate the Gaussian
integration in the derivatives of the mass functions 
in  Eqs.~(\ref{eq:primeP}) and ~(\ref{eq:primeV}).
The exponent has the form $ak^2+2kr+z_0$, where $r=b\,p$. 
Next, with the help of the properties ($k$ is the loop momentum)
\begin{equation}
\left.
\begin{aligned}
	k^\mu\, \exp(ak^2+2kr+z_0) &=\frac{1}{2}\frac{\partial }
	{\partial r_\mu }\exp(ak^2+2kr+z_0)\\
	k^\mu k^\nu\, \exp(ak^2+2kr+z_0) &=
	\frac{1}{2}\frac{\partial }{\partial r_\mu } 
	\frac{1}{2} \frac{\partial }{\partial r_\nu }         \exp(ak^2+2kr+z_0)
	\\
	\text{etc.}&
\end{aligned}
\right\},
\label{eq:change-to-r}
\end{equation}
we replace
$\not\! k $ by $ {\not\! \partial}_r 
= \gamma^\mu\frac{\partial}{\partial r_\mu}$,
which allows us to exchange the tensor integrations
for a differentiation of the Gaussian exponent. For example,
Eq.~(\ref{eq:massP}) now has the form
\begin{equation}
\widetilde\Pi_P(p^2) = \frac{3g^2_P}{16\pi^2}
\int\limits_0^\infty\!\! \int\limits_0^\infty\!
\frac{d\alpha_1 d\alpha_2}{a^2} 
\,
\Tr\left[\gamma^5 (m_1+{\not\! \partial}_r+w_1\not\!p)\gamma^5 
(m_2+{\not\! \partial}_r-w_2\not\!p)\right]
e^{-\frac{r^2}{a}+z_0}.
\label{eq:Pmass2}\,
\end{equation}
The $r$-dependent Gaussian exponent $e^{-r^2/a}$ is moved to the left through the differential operator $\not\! \partial_r$ by using the properties
\begin{eqnarray}
	\frac{\partial}{\partial r_\mu}\,e^{-r^2/a} &=& e^{-r^2/a}
	\left[-\frac{2r^\mu}{a}+\frac{\partial}{\partial r_\mu}\right],
	\nn[1.2ex]
	\frac{\partial}{\partial r_\mu}\,
	\frac{\partial}{\partial r_\nu}\,e^{-r^2/a} &=& e^{-r^2/a}
	\left[-\frac{2r^\mu}{a}+\frac{\partial}{\partial r_\mu}\right]\cdot
	\left[-\frac{2r^\nu}{a}+\frac{\partial}{\partial r_\nu}\right],
	\nn[1.2ex]
	\text{etc.}&&
	\label{eq:dif}
\end{eqnarray}
Finally, the derivatives are moved to the right by using
the commutation relation
\begin{equation}
\left[\frac{\partial}{\partial r_\mu},r^\nu \right]
=g^{\mu\nu}.
\label{eq:comrel}
\end{equation}
The last step is done by using a \textsc{form} code that
works for any numbers of loops and propagators.

The remaining integrals over the Fock-Schwinger parameters 
$0\le \alpha_i<\infty$ are treated by introducing an additional integration that converts the set of 
these parameters into a simplex as follows:
\begin{equation}
	\prod\limits_{i=1}^n\int\limits_0^{\infty} 
	\!\! d\alpha_i f(\alpha_1,\ldots,\alpha_n)
	=\int\limits_0^{\infty} \!\! dtt^{n-1}
	\prod\limits_{i=1}^n \int\!\!d\alpha_i 
	\delta\left(1-\sum\limits_{i=1}^n\alpha_i\right)
	f(t\alpha_1,\ldots,t\alpha_n).
	\label{eq:simplex}  
\end{equation}
As an example, we show below the expression for the derivative of the mass function of a pseudoscalar meson:
\begin{eqnarray}
	\widetilde\Pi'_P(p^2) &=& \frac{3g^2_P}{4\pi^2}\int\limits_0^{\infty} \!\! 
	\frac{dt\,t}{a_P^2} \int\limits_0^1\!\!d\alpha\,
	e^{-t\,z_0 + z_P}\,f_P(t,\alpha),
	\label{eq:prime_fin}\\[2ex]
	z_0 &=&  \alpha m^2_{q_1} +(1-\alpha)m^2_{q_2} - \alpha(1-\alpha) p^2,
	\nn 
	z_P &=& \frac{2s_Pt}{a_P} (\alpha-w_2)^2 p^2,
	\nn
	a_P &=& 2s_P+t , \qquad s_P=1/\Lambda^2_P. 
	\nonumber
\end{eqnarray}
The function $f_P(t,\alpha)$ arises from trace evaluation.

At this stage, an infrared cutoff is introduced to avoid any possible thresholds in the Feynman diagram:
\begin{equation}
	\int\limits_0^\infty dt (\ldots) \to \int\limits_0^{1/\lambda^2} dt (\ldots).
	\label{eq:conf}
\end{equation}
The infrared cutoff parameter $\lambda$ effectively guarantees the confinement of quarks within hadrons. This method is generic and can be used for diagrams with an arbitrary 
number of loops and propagators. In the CCQM, the parameter $\lambda$ is assumed to be universal. The numerical evaluation of the final integral in Eq.~(\ref{eq:conf}) was done by \textsc{FORTRAN} code using the NAG library.

Let us enumerate the number of free parameters in the CCQM and describe the fitting process to determine their values. For a given meson $M_i$, five parameters are needed: the size parameter $\Lambda_{M_i}$,  meson-quark coupling $g_{M_i}$, two of the four constituent quark masses $m_{q_j}$, ($m_u=m_d, m_c, m_s, m_b$), and universal cutoff parameter $\lambda$. If $n_M$ mesons are considered, one has $2n_M+5$ model parameters. However, the compositeness condition in Eq.~(\ref{eq:compositeness}) provides $n_M$ constraints. These constraints are used to exclude the coupling $g_{M_i}$ from the parameter set. The remaining $n_M+5$ parameters are obtained by fitting experimental data. Naturally, we first chose experimental values for the leptonic decay constants of $n_M$ mesons to fit the model parameters. Whenever experimental data were not available, or were available but with large errors, we decided to use LQCD results instead. 

\begin{figure}[htbp]
	\includegraphics[scale=0.45]{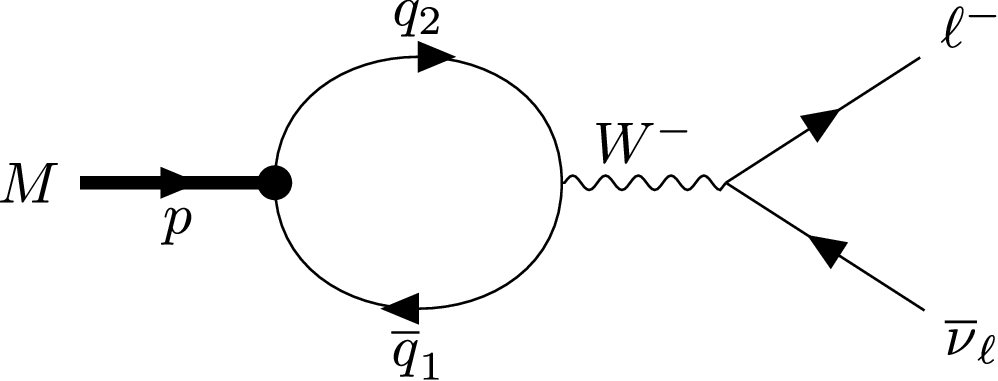}
	\caption{Quark model diagram for meson leptonic decay.}
	\label{fig:leptonic}
\end{figure}

The matrix elements of the leptonic decays are described by
the Feynman diagram shown in Fig.~\ref{fig:leptonic}.
The leptonic decay constants of the pseudoscalar and vector mesons are defined in the CCQM as
\begin{eqnarray}
	N_c\, g_P\! \int\!\! \frac{d^4k}{ (2\pi)^4 i}\, \widetilde\Phi_P(-k^2)\,
	{\rm tr} \biggl[O^{\,\mu} S_1(k+w_1 p) \gamma^5 S_2(k-w_2 p) \biggr] 
	&=&f_P p^\mu ,
	\nn
	N_c\, g_V\! \int\!\! \frac{d^4k}{ (2\pi)^4 i}\, \widetilde\Phi_V(-k^2)\,
	{\rm tr} \biggl[O^{\,\mu} S_1(k+w_1 p)\not\!\epsilon_V  S_2(k-w_2 p) \biggr] 
	&=& m_V f_V \epsilon_V^\mu,
	\label{eq:lept}
\end{eqnarray}
where $N_c=3$ is the number of colors  and $O^\mu=\gamma^\mu(1-\gamma_5)$ is the weak Dirac matrix with left chirality. The mesons are taken on their mass-shells. The matrix elements in Eq.~(\ref{eq:lept}) are calculated in a similar manner as that for the case of the mass functions described above.

In addition, we decided to use eight fundamental electromagnetic decays to establish further constraints on the model parameters. These decays are listed in Table~\ref{tab:fit-elmag}. The results of (overconstrained) least-squares fitting for the leptonic decay constants and electromagnetic decay widths are reported in Tables~\ref{tab:fit-leptonic} and \ref{tab:fit-elmag}. The results of the fitting for the values of quark masses, size parameters, and  infrared cutoff parameter are reported in Tables~\ref{tab:size_parameter} and \ref{tab:quark_mass}.

\begin{table}[ht]
	\caption{Input values for the leptonic decay constants $f_M$ (in MeV) and our least-squares fitting values~\cite{Ivanov:2011aa}}
	\label{tab:fit-leptonic}
	\begin{center}
		\begin{tabular}{c|c|c|c||c|c|c|c}
\hline\hline
& Fit Values & Input & Ref. & & Fit Values & Input & Ref.\\
\hline
$f_\pi$ & 128.7(6.4) & $130.4\pm 0.2$ & \cite{ParticleDataGroup:2010dbb, Rosner:2010ak} & $f_\omega$ & 198.5(9.9) & $198\pm 2$ & \cite{ParticleDataGroup:2010dbb}\\
$f_K$ & 156.1(7.8) & $156.1\pm 0.8$ & \cite{ParticleDataGroup:2010dbb, Rosner:2010ak} & $f_\phi$ & 228.2(11.4) & $227\pm 2$ & \cite{ParticleDataGroup:2010dbb}\\
$f_D$ & 205.9(10.3) & $206.7\pm 8.9$ & \cite{ParticleDataGroup:2010dbb, Rosner:2010ak} & $f_{J/\psi}$ & 415.0(20.8) & $415\pm 7$ & \cite{ParticleDataGroup:2010dbb}\\
$f_{D_s}$ & 257.5(12.9) & $257.5\pm 6.1$ & \cite{ParticleDataGroup:2010dbb, Rosner:2010ak} & $f_{K^*}$ & 213.7(10.7) & $217\pm 7$ & \cite{ParticleDataGroup:2010dbb}\\
$f_B$ & 191.1(9.6) & $192.8\pm 9.9$ & \cite{Laiho:2009eu} & $f_{D^*}$ & 243.3(12.2) & $245\pm 20$ & \cite{Becirevic:1998ua}\\
$f_{B_s}$ & 234.9(11.7) & $238.8\pm 9.5$ & \cite{Laiho:2009eu} & $f_{D_s^*}$ & 272.0(13.6) & $272\pm 26$ & \cite{Becirevic:1998ua}\\
$f_{B_c}$ & 489.0(24.5) & $489\pm 5$ & \cite{Chiu:2007km} & $f_{B^*}$ & 196.0(9.8) & $196\pm 44$ & \cite{Becirevic:1998ua}\\
$f_\rho$ & 221.1(11.1) & $221\pm 1$ & \cite{ParticleDataGroup:2010dbb} & $f_{B_s^*}$ & 229.0(11.5) & $229\pm 46$ & \cite{Becirevic:1998ua}
			\\			
			\hline\hline
		\end{tabular}
	\end{center}
\end{table} 

\begin{table}[ht]
	\caption{Input values for some basic electromagnetic decay widths and our least-squares fitting values (in keV)~\cite{Ivanov:2011aa}.}
	\label{tab:fit-elmag}
	\begin{center}
		\begin{tabular}{c|c|c}
			\hline\hline
Process	& Fit Values & Input~\cite{ParticleDataGroup:2010dbb}\\
\hline
$\pi^0\to \gamma\gamma$	& $5.06(25)\times 10^{-3}$ & $(7.7\pm 0.4)\times 10^{-3}$\\	
$\eta_c\to \gamma\gamma$& 1.61(8) & $1.8\pm 0.8$\\
$\rho^+\to \pi^+\gamma$	& 76.0(3.8) & $67\pm 7$\\
$\omega\to \pi^0\gamma$	& 672(34) & $703\pm 25$\\
$K^{*+}\to K^+\gamma$	& 55.1(2.8) & $50\pm 5$\\
$K^{*0}\to K^0\gamma$	& 116(6) & $116\pm 10$\\
$D^{*+}\to D^+\gamma$	& 1.22(6) & $1.5\pm 0.5$\\
$J/\psi\to \eta_c\gamma$& 1.43(7) & $1.58\pm 0.37$\\
\hline\hline
\end{tabular}
\end{center}
\end{table}

Once the free parameters are fixed, the CCQM can be employed as a frame-independent tool for hadronic calculation. Note that this data fitting was performed in 2011 (see Ref.~\cite{Ivanov:2011aa}). One of the advantages of the CCQM is the possibility to calculate hadronic quantities in the full physical range of momentum transfer without any extrapolation. Another advantage of the model is that it can be used to treat not only mesons~\cite{Faessler:2002ut, Ivanov:2006ni, Ivanov:2020iad, Ivanov:2019nqd}, but also baryons~\cite{Gutsche:2013pp, Gutsche:2018nks, Groote:2021ayy}, tetraquarks~\cite{Dubnicka:2011mm, Goerke:2016hxf}, and other multiquark states~\cite{Gutsche:2017twh, Soni:2020sgn} in a consistent way. 

Concerning the estimation of theoretical errors in the CCQM, we use MINUIT for fitting, which is based on 	$\chi^2$ minimalization. When we performed the fitting, we obtained the best-fitting values for model parameters, which are listed in  Tables~\ref{tab:size_parameter} and \ref{tab:quark_mass}. We did not consider the errors of these parameters because it is complicated to take into account the error propagations in the final physical  predictions. Instead, we followed an easier but less rigorous approach for error estimation. We only used the best-fitting parameters for further calculation. Then, we observed that the fitted values deviated from experimental data by approximately $5\%-10\%$. In our model, hadronic quantities are calculated in a similar manner to the calculations of the leptonic and electromagnetic decay constants. Therefore, we estimated the errors of the hadronic quantities to be approximately $5\%-10\%$. When these hadronic quantities are used for further calculations, for example, the decay widths, the errors accumulate. Consequently, we estimated the error for the decay widths to be approximately $10\%-20\%$. In particular, in this study, we estimated the error for the couplings $g_{D*D\gamma}$ to be $10\%$ and the error for the decay widths to be $20\%$.

\begin{table}[ht]
	\caption{Meson size parameters (in GeV)~\cite{Ivanov:2011aa}.}\label{tab:size_parameter}
	\renewcommand{\arraystretch}{0.7}
	\begin{ruledtabular}
		\begin{tabular}{cccccccc}
		$\Lambda_{\pi}$ & $\Lambda_K$ &	$\Lambda_{D}$ & $\Lambda_{D_s}$ & $\Lambda_B$ & $\Lambda_{B_s}$ & $\Lambda_{B_c}$ & $\Lambda_{\rho}$\\
			\hline
		0.87(10) & 1.04(26) &	1.47(39) & 1.57(42) & 1.88(26) & 1.95(27) & 2.42(10) & 0.61(7) \\
		\hline\hline 
		$\Lambda_{\omega}$ & $\Lambda_{\phi}$ &	$\Lambda_{J/\psi}$ & $\Lambda_{K^*}$ & $\Lambda_{D^*}$ & $\Lambda_{D^*_s}$ & $\Lambda_{B^*}$ & $\Lambda_{B_s^*}$\\
		0.47(5) & 0.88(10) & 1.48(17) & 0.72(18) & 1.16(22) & 1.17(22) & 1.72(27) & 1.71(27) \\
		\end{tabular}
	\end{ruledtabular}
\end{table}
\begin{table}[ht]
	\caption{Quark masses and infrared cutoff parameter (in GeV)~\cite{Ivanov:2011aa}.}\label{tab:quark_mass}
	\renewcommand{\arraystretch}{0.7}
	\begin{ruledtabular}
		\begin{tabular}{ccccc}
			$m_{u/d}$ & $m_s$ &  $m_c$ &  $m_b$ & $\lambda$ \\
			\hline
			0.235(12) & 0.424(19) & 2.160(71) & 5.090(21) & 0.181(9)
		\end{tabular}
	\end{ruledtabular}
\end{table}

\section{Radiative decays \bm{$D^*_{(s)}\to D_{(s)}\gamma}$}
\label{sec:decay}
\begin{figure}[ht]
	\begin{center}
		\begin{tabular}{cc} 
			\includegraphics[width=0.5\textwidth]{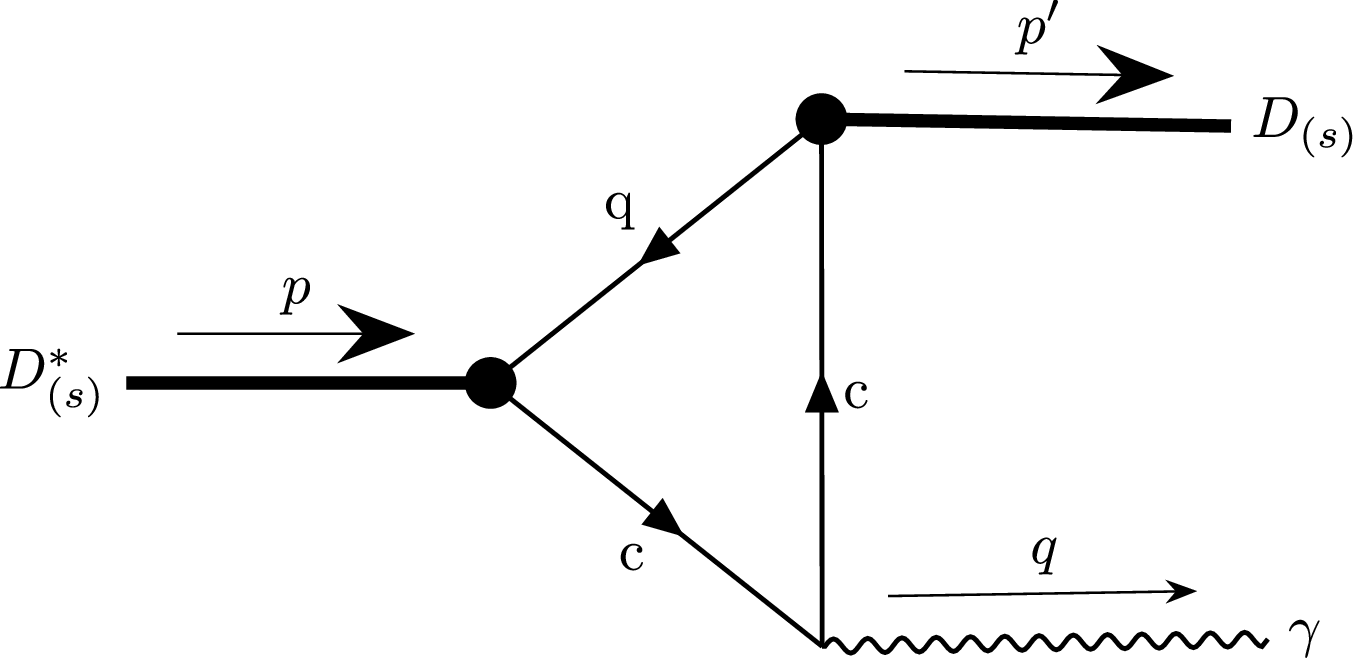} &
			\includegraphics[width=0.5\textwidth]{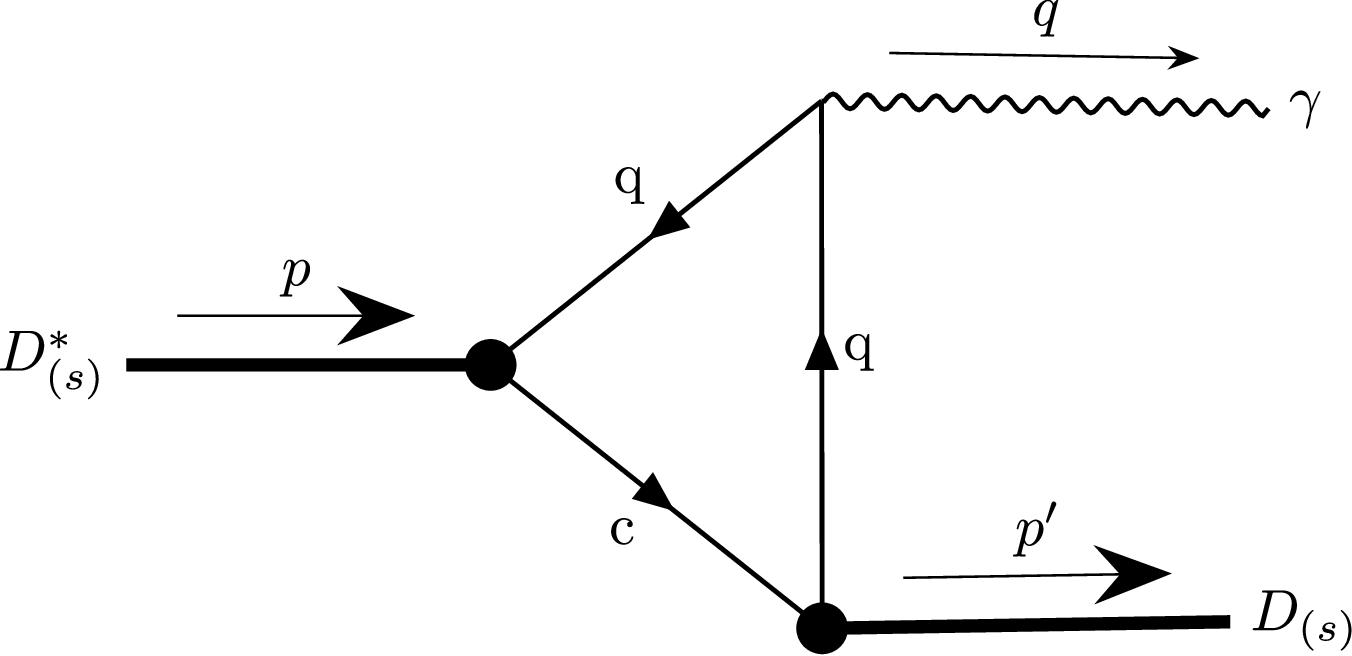} \end{tabular}
	\end{center}
	\caption{\label{fig:VPg}
		Feynman diagrams for radiative decays $D^*_{(s)}\to D_{(s)}\gamma$.}
\end{figure}

The radiative decays $D^*_{(s)}\to D_{(s)}\gamma$ are described by the Feynman diagrams shown in Fig.~\ref{fig:VPg}. The coupling between the quarks and the photon is described by the interaction Lagrangian,
\begin{equation}
	\mathcal{L}_{\mathrm{int}}^{\mathrm{em}}(x) = eA_\mu(x)J_{\mathrm{em}}^\mu(x),\qquad J_{\mathrm{em}}^\mu(x) = e_c\bar{c}(x)\gamma^\mu c(x)
	+ e_q\bar{q}(x)\gamma^\mu q(x),
\end{equation}
where $e_c$ and $e_q$ are the quark charges in units of $e$.

The transition amplitude is expressed as
\begin{eqnarray}
\langle \vec{p^\prime};\vec{q},\epsilon_\mu^\gamma|S-I|\vec{p},\epsilon_\nu^{D^*}\rangle &=&(i)^3 e g_{D^*} g_D \epsilon^{D^*}_\nu(p)\epsilon^\gamma_\mu(q)\nonumber\\
&&\times\iiint dx\,dy\,dz\,
e^{-ipx+ip^\prime y+iqz} \langle\mathrm{T}{\bar{J}_{D^*}^\nu(x) J_{\mathrm{em}}^\mu(z) J_D(y)}\rangle_0\nonumber\\
&=& (2\pi)^4 i \delta(p-p^\prime-q)\mathcal{M}(D^*\to D\gamma),
\end{eqnarray}
where
\begin{eqnarray}
\mathcal{M}(D^*\to D\gamma) &=& (-3i)e g_{D^*} g_D \epsilon^{D^*}_\nu(p)\epsilon^\gamma_\mu(q) (e_c \mathcal{M}^{\mu\nu}_c + e_q \mathcal{M}^{\mu\nu}_q),\\
\mathcal{M}^{\mu\nu}_c &=& \int \frac{dk}{(2\pi)^4i}\widetilde{\Phi}_{D^*}\big[-(k-\omega_2p)^2\big]\widetilde{\Phi}_{D}\big[-(k-\omega_2p^\prime)^2\big]\nonumber\\
&&\times\mathrm{tr}\big[S_q(k)\gamma^\nu S_c(k-p)\gamma^\mu S_c(k-p^\prime)\gamma^5\big],\\
\mathcal{M}^{\mu\nu}_q &=& \int \frac{dk}{(2\pi)^4i}\widetilde{\Phi}_{D^*}\big[-(k+\omega_1p)^2\big]\widetilde{\Phi}_{D}\big[-(k+\omega_1p^\prime)^2\big]\nonumber\\
&&\times\mathrm{tr}\big[S_q(k+p^\prime)\gamma^\mu S_q(k+p)\gamma^\nu S_c(k)\gamma^5\big].
\end{eqnarray}
The ratios of quark masses now read $\omega_1=m_{c}/(m_{c}+m_{q})$ and $\omega_2=m_{q}/(m_{c}+m_{q})$ with $q=u,d,s$.

To calculate the matrix element $\mathcal{M}(D^*\to D\gamma)$, the Gaussian form is substituted for the vertex functions in Eq.~(\ref{eq:vertexf}) and Fock-Schwinger representation of quark propagators in Eg.~(\ref{eq:Fock}), and then the techniques described by Eqs.~(\ref{eq:Wick})--(\ref{eq:conf}) are applied.

Next, using the transversality conditions $\epsilon_\mu^\gamma(q) q^\mu = 0$ and $\epsilon_\nu^{D^*}(p) p^\nu = 0$, the matrix element can be expressed in the form
\begin{equation}
	\mathcal{M}(D^*\to D\gamma)=e\, g_{D^*D\gamma}\, \varepsilon^{\alpha\beta\mu\nu}\,p_\alpha\, q_\beta\,\epsilon_\mu^\gamma(q)\,\epsilon_\nu^{D^*}(p),
\end{equation}
where $g_{D^*D\gamma}=e_c\,I_c(m^2_{D^*},m^2_D)+e_q\,I_q(m^2_{D^*},m^2_D)$ is the radiative decay constant. The quantities $I_{c(q)}(m^2_{D^*},m^2_D)$ are threefold integrals which are calculated numerically. 
The expression for $I_{c}(m^2_{D^*},m^2_D)$ reads
\begin{eqnarray}
	\label{eq:I}
	I_{c}(m^2_{D^*},m^2_D) &=& g_D g_{D^*} \frac{N_c}{4\pi^2}\int\limits_0^{1/{\lambda^2}} \!\! 
	\frac{dt\,t^2}{(s+t)^2} \int\limits\!\!d\alpha^3 \delta\big(1-\sum\limits_{i=1}^3\alpha_i \big)\\ \nonumber
	&&\times \left[m_c\omega_2+m_q\omega_1+\frac{t}{s+t}(m_c-m_q)(\omega_1-\alpha_2) \right]\exp\left(-tz_0 + \frac{st}{s+t}z_1\right),
	\\[2ex] \nonumber
	z_0 &=&  (1-\alpha_2)m_c^2+\alpha_2m_q^2-\alpha_1\alpha_2 m_{D^*}^2-\alpha_2\alpha_3 m_D^2,
	\nn 
	z_1 &=& m_{D^*}^2\left(\alpha_1-\omega_2\frac{s_{D^*}}{s}\right)(\omega_1-\alpha_2)+m_D^2(\alpha_2-\omega_1)\left(\alpha_1+\alpha_2-\frac{s_{D^*}}{s}-\omega_1\frac{s_{D}}{s}\right),
	\nn
	s &=& s_{D^*}+s_D , \qquad s_{D^{(*)}}=1/\Lambda^2_{D^{(*)}}. 
	\nonumber
\end{eqnarray}	
The expression for $I_{q}(m^2_{D^*},m^2_D)$ can be obtained by simply exchanging $m_c \leftrightarrow m_q$ and $\omega_1 \leftrightarrow \omega_2$, i.e. $I_{q}(m^2_{D^*},m^2_D,m_q, m_c, \omega_1,\omega_2)=I_{c}(m^2_{D^*},m^2_D,m_c, m_q, \omega_2,\omega_1)$.

Finally, the decay width for $D^*\to D\gamma$ is given by
\begin{equation}
	\label{eqn:width}
	\Gamma(D^*\to D\gamma) = \frac{\alpha}{24}  m_{D^*}^3\left(1-
	\frac{m_{D}^2}{m_{D^*}^2}\right)^3g_{D^*D\gamma}^2.
\end{equation}

\section{Numerical results and discussion}
\label{sec:result}

Our results for the radiative decay constants are listed in Table~\ref{tab:gpvg} together with results obtained in other theoretical approaches. 
	It is worth mentioning that the sign difference in the predictions in Table~\ref{tab:gpvg} is simply due to the choice when one defines the matrix element and does not affect the physical observables, i.e., the decay width, given that the last is proportional to the square of the matrix element. In other words, the sign of the radiative decay constants cannot be measured experimentally.
The experimental value $g_{D^{*+}D^+\gamma}=-0.47(7)\,\textrm{GeV}^{-1}$ was extracted from the total width of $D^{*+}$ and measured branching fraction $\mathcal{B}(D^{*+}\to D^+\gamma)$~\cite{Pullin:2021ebn}. Meanwhile, the total width of $D^{*0}$ is still unknown experimentally. However, it can be calculated using the isospin symmetry related to the strong couplings $g_{D^{*+}D\pi}$ and $g_{D^{*0}D\pi}$~\cite{Becirevic:2012pf, Becirevic:2009xp}, which yields $\Gamma(D^{*0})=56.5(14.0)\,\textrm{keV}$~\cite{Pullin:2021ebn}. Then, the value $g_{D^{*0}D^0\gamma}=1.77(16)\,\textrm{GeV}^{-1}$ was obtained with the help of the measured $\mathcal{B}(D^{*0}\to D^{0}\gamma)$~\cite{Pullin:2021ebn}. Note that our predictions $g_{D^{*+}D^+\gamma}=-0.45(9)\,\textrm{GeV}^{-1}$ and $g_{D^{*0}D^0\gamma}=1.72(34)\,\textrm{GeV}^{-1}$ are in full agreement with the current experimental data. Note also that the decay width of the channel $D^{*+}\to D^+\gamma$ was included in the model parameter fitting (see Ref.~\cite{Ivanov:2011aa} and Table~\ref{tab:fit-elmag}). However, channels $D^{*0}\to D^0\gamma$ and $D^{*+}_s\to D_s^+\gamma$ were not included in the fit.

\begin{table}[ht]
	\caption{Radiative decay constants in the CCQM and other approaches (all in $\textrm{GeV}^{-1}$)}
	\label{tab:gpvg}
	\begin{center}
		\begin{tabular}{c|c|c|c}
			\hline\hline
			Ref.  & $g_{D^{*+}D^+\gamma}$  &  $g_{D^{*0}D^0\gamma}$ & $g_{D^{*+}_sD^+_s\gamma}$ \\
			\hline
			This work              &   $-0.45(9)$    &    $1.72(34)$      &  $-0.29(6)$ \\
			LCSR (NLO)~\cite{Pullin:2021ebn}  & $0.40^{+0.12}_{-0.13}$ & $-2.11^{+0.35}_{-0.34}$ & $0.60^{+0.19}_{-0.18}$ \\
			LCSR (NLO)\cite{Li:2020rcg}      & $-0.15^{+0.11}_{-0.10}$ & $1.48^{+0.29}_{-0.27}$ & $-0.079^{+0.086}_{-0.078}$  \\
			HQET$+$CQM~\cite{Cheung:2014cka}    & $-0.38^{+0.05}_{-0.04}$ & $1.91^{+0.09}_{-0.09}$ & --\\
			LQCD~\cite{Becirevic:2009xp}$^\dagger$,~\cite{Donald:2013sra}$^*$,~\cite{Guadagnoli:2023zym}$^{**}$     & $-0.2(3)^\dagger $ & $2.0(6)^\dagger $  &  $0.11(2)^*$, $0.04(1)^{**}$ \\
			Bag Model~\cite{Orsland:1998de}  & 0.5 & 1.1 & --\\
			RQM~\cite{Jaus:1996np}  & $-0.30$ & 1.85 & --\\		
			RQM~\cite{Goity:2000dk}   & $-0.44(6)$ & $2.15(11) $ & $-0.19(3)$\\		
			LCSR~\cite{Aliev:1995zlh} & $-0.50(12)$ & $1.52(25)$ & -- \\			
			QCDSR~\cite{Aliev:1994nq} & $-0.19^{+0.03}_{-0.02}$ & $0.62^{+0.03}_{-0.03}$ & $-0.20^{+0.03}_{-0.03}$ \\				
			HQET$+$VMD~\cite{Colangelo:1993zq} & $-0.29^{+0.19}_{-0.11}$ & $1.60^{+0.35}_{-0.45}$ & $-0.19^{+0.19}_{-0.08}$ \\			
			HH$\chi$PT~\cite{Amundson:1992yp} & $-0.27(5)$ & $2.19(11)$ & $0.041(56)$  \\			
			Experiment~\cite{Workman:2022ynf} & $-0.47(7)$ & $1.77(16)$ & -- \\			
			\hline\hline
		\end{tabular}
	\end{center}
\end{table} 

In the case of $g_{D^{*+}_sD^+_s\gamma}$, the theoretical predictions vary significantly with respect to the case of $g_{D^{*+}D^+\gamma}$ and $g_{D^{*0}D^0\gamma}$. 
The central value of our result, $|g_{D^{*+}_sD^+_s\gamma}|=0.29(6)\,\textrm{GeV}^{-1}$, is larger than all other theoretical predictions, except for a recent value, $|g_{D^{*+}_sD^+_s\gamma}|=0.60(19)\,\textrm{GeV}^{-1}$, obtained by using LCSR(NLO)~\cite{Pullin:2021ebn}. 
Our result deviates from others within $2\sigma$ owing to large uncertainties in the predictions.
In Ref.~\cite{Pullin:2021ebn}, a discussion on $D$-spin symmetry for the decays $D^{*+}\to D^+\gamma$ was also provided. 
Here, the $D$-spin symmetry	is understood as the symmetry under the exchange of down and strange quarks $d\leftrightarrow s$. As pointed out in Ref.~\cite{Pullin:2021ebn}, this is a good approximate symmetry in QED. Given that the decay $D_s^{*+}\to D_s^+\gamma$ can be obtained from the decay $D^{*+}\to D^+\gamma$ by exchanging $d\leftrightarrow s$, the order of this symmetry breaking is expected not to be too far from that in QED.
It was roughly estimated that the deviation between $g_{D^{*+}D^+\gamma}$ and $g_{D^{*+}_sD^+_s\gamma}$ is approximately $20$--$30\%$. Our results show a deviation of approximately $35\%$ which is reasonable. 
Note also from Table~\ref{tab:gpvg} the difference between the predictions of Refs.~\cite{Pullin:2021ebn} and \cite{Li:2020rcg}, despite the similarity in the approach, which is LCSR at NLO. The authors of Ref.~\cite{Pullin:2021ebn} explained this issue as the result of computational differences between both studies. In particular, the twist-1 $\mathcal{O}(\alpha_s)$-corrections were calculated in Ref.~\cite{Pullin:2021ebn} while they were neglected in Ref.~\cite{Li:2020rcg}, and the twist-4 corrections were dismissed in Ref.~\cite{Pullin:2021ebn} while they were considered in Ref.~\cite{Li:2020rcg}.     
\begin{table}[ht]
	\caption{Decay widths $\Gamma(D^*_{(s)}\to D_{(s)}\gamma)$ in the CCQM and other approaches (all in $\textrm{keV}$)}
	\label{tab:width_VPg}
	\begin{center}
		\begin{tabular}{c|c|c|c}
			\hline\hline
			Ref.  & $\Gamma(D^{*+}\to D^+\gamma)$  &  $\Gamma(D^{*0}\to D^0\gamma)$ & $\Gamma(D^{*+}_s\to D^+_s\gamma)$\\
			\hline
			This work  &   $1.21(48)$    &    $18.4(7.4)$      &  $0.55(22)$\\
			LCSR (NLO)~\cite{Pullin:2021ebn}  & $0.96^{+0.58}_{-0.62}$ & $27.83^{+9.23}_{-9.50}$ & $2.36^{+1.49}_{-1.41}$ \\
			LQCD~\cite{Becirevic:2009xp}$^\dagger$,~\cite{Donald:2013sra}$^*$     & $0.8(7)^\dagger $ & $27(14)^\dagger $  &  $0.066(26)^*$ \\
			HQET$+$CQM~\cite{Cheung:2014cka}    & $0.9\pm 0.3$ & $22.7\pm 2.2$ & --  \\
			Bag Model~\cite{Orsland:1998de}  & 1.73 & 7.18 & --\\
				NJL Model~\cite{Deng:2013uca}
			 & 0.7 & 19.4 & 0.09
			 \\
			RQM~\cite{Colangelo:1994jc} & 0.46 & 23.05  & 0.38 \\
			RQM~\cite{Jaus:1996np}  & 0.56  & 21.69  & --\\
			RQM~\cite{Goity:2000dk}   & (0.94--1.42) & (26.0--32.0) & (0.2--0.3)\\
			RQM~\cite{Ebert:2002xz} & 1.04 & 11.5 & 0.19\\
			LFQM~\cite{Choi:2007se}  & $0.90(2)$ & $20.0(3)$ & $0.18(1)$\\		
			QCDSR~\cite{Zhu:1996qy} & $0.23(10)$ & $12.9(2.0)$ & $0.13(5)$\\
			LCSR~\cite{Aliev:1995zlh} & 1.50 & 14.40 & -- \\
			QCDSR~\cite{Aliev:1994nq} & $0.22(6)$ & $2.43(21)$ & $0.25(8)$\\				
			HQET$+$VMD~\cite{Colangelo:1993zq} & $0.51(57)$ & $16.0(10.8)$ & $0.24(24)$   \\
			Experiment~\cite{Workman:2022ynf} & $1.33(36)$ & $19.9(5.5)$ & -- \\
			\hline\hline
		\end{tabular}
	\end{center}
\end{table}

Table~\ref{tab:width_VPg} compares the predictions for the width of the $D^*_{(s)}$ radiative decays. Available theoretical predictions for $\Gamma(D^{*+}\to D^+\gamma)$ vary in the range $(0.2 - 1.7)$ keV. Our result $\Gamma(D^{*+}\to D^+\gamma)=1.21(48)$ agrees well with experimental data, suggesting the width of the decay $D^{*+}\to D^+\gamma$ to be at the higher end of the predicted range. Note that the experimental value $\Gamma(D^{*+}\to D^+\gamma)=1.33(36)\,\textrm{keV}$ was obtained from the world average of the total width $\Gamma(D^{*+})=83.4(1.8)\,\textrm{keV}$ and  branching fraction $\mathcal{B}(D^{*+}\to D^+\gamma)=0.016(4)$ reported by Particle Data Group (PDG)~\cite{Workman:2022ynf}. 
	We stress again that the decay $D^{*+}\to D^+\gamma$ was included in our model parameter fitting (see Table~\ref{tab:fit-elmag}). However, this fitting was performed in 2011~\cite{Ivanov:2011aa}, and the input value $\Gamma(D^{*+}\to D^+\gamma)=1.5(5)\,\textrm{keV}$ was taken from PDG at that time~\cite{ParticleDataGroup:2010dbb}. The slight difference for the central values of $\Gamma(D^{*+}\to D^+\gamma)$ given in Tables~\ref{tab:width_VPg} and \ref{tab:fit-elmag}, which are 1.21 and 1.22, respectively, is because we used the most updated values for meson masses in this study.

The experimental value $\Gamma(D^{*0}\to D^0\gamma)=19.9(5.5)\,\textrm{keV}$ in Table~\ref{tab:width_VPg} was calculated by using the estimation $\Gamma(D^{*0})=56.5(14.0)\,\textrm{keV}$ mentioned above and the branching fraction $\mathcal{B}(D^{*0}\to D^0\gamma)=0.353(9)$ fitted by PDG~\cite{Workman:2022ynf}. Our prediction $\Gamma(D^{*0}\to D^0\gamma)=18.4(7.4)\,\textrm{keV}$ also agrees  well with this experimental value.

At this point, the ratio of decay widths $\Gamma(D^{*+}\to D^+\gamma)/\Gamma(D^{*0}\to D^0\gamma)$ can be checked. It follows from Eq.~(\ref{eqn:width}) that
\begin{eqnarray}
	\frac{\Gamma(D^{*+}\to D^+\gamma)}{\Gamma(D^{*0}\to D^0\gamma)} &=& \frac{m^3_{D^{*0}}}{m^3_{D^{*+}}} \left(\frac{m^2_{D^{*+}}-m^2_{D^+}}{m^2_{D^{*0}}-m^2_{D^0}}\right)^3
	\left[\frac{e_c I_c(m^2_{D^{*+}},m^2_{D^+})+e_d I_d(m^2_{D^{*+}},m^2_{D^+})}{e_c I_c(m^2_{D^{*0}},m^2_{D^0})+e_u I_u(m^2_{D^{*0}},m^2_{D^0})}\right]^2\nonumber\\
	&\approx& \frac{1}{4}\left[\frac{2I_c(m^2_{D^{*+}},m^2_{D^+})-I_d(m^2_{D^{*+}},m^2_{D^+})}{I_c(m^2_{D^{*+}},m^2_{D^+})+I_u(m^2_{D^{*+}},m^2_{D^+})}\right]^2.
\end{eqnarray}
At the final step, we have neglected the mass differences between charged and neutral $D^{(*)}$ mesons. Note also that we have assumed from the beginning that $m_d=m_u$.
In the heavy quark limit (HQL) $m_c\to\infty$, the integral $I_c$ is suppressed as $1/m_c$ and the width ratio yields $0.25$. However, from Table~\ref{tab:width_VPg}, one obtains $\Gamma(D^{*+}\to D^+\gamma)/\Gamma(D^{*0}\to D^0\gamma)=0.066$. Therefore, the use of HQL is not suitable for the decays $D^{*}\to D\gamma$. This is opposed to the case of the decays $B^{*}\to B\gamma$, in which the use of HQL for $b$ quark is reliable, as pointed out in a previous paper of us~\cite{Ivanov:2022nnq}.

In the case of the decay $D^*_s\to D_s\gamma$, there is still much speculation about its width. Predictions for $\Gamma(D^*_s\to D_s\gamma)$ range from 0.07 keV to 2.4 keV. 	Our result $\Gamma(D^*_s\to D_s\gamma)=0.55(22)\,\textrm{keV}$ deviates from the recent prediction based on LCSR at NLO $\Gamma(D^*_s\to D_s\gamma)=2.36^{+1.49}_{-1.41}\,\textrm{keV}$~\cite{Pullin:2021ebn} and the first LQCD result $\Gamma(D^*_s\to D_s\gamma)=0.066(26)\,\textrm{keV}$~\cite{Donald:2013sra} by approximately $2\sigma$. 	Note also that the central value of our result $\Gamma(D^*_s\to D_s\gamma)=0.55(22)\,\textrm{keV}$ is larger than that of most predictions; in particular, it exceeds the LQCD central value $\Gamma(D^*_s\to D_s\gamma)=0.066\,\textrm{keV}$~\cite{Donald:2013sra} by roughly an order of magnitude. By contrast, the recent prediction based on LCSR at NLO $\Gamma(D^*_s\to D_s\gamma)=2.36^{+1.49}_{-1.41}\,\textrm{keV}$~\cite{Pullin:2021ebn} has a central value approximately four times as large as our value. Other theoretical studies seem to agree on the central value $\Gamma(D^*_s\to D_s\gamma)\approx 0.2\,\textrm{keV}$.

It is instructive to estimate the width of the decay $D^*_s\to D_s\gamma$ using the recent measurement of the leptonic decay $D_s^{*+}\to e^+\nu_e$ performed by the BESIII collaboration~\cite{BESIII:2023zjq}. This is the first measurement of the decay branching fraction; it reads $\mathcal{B}(D_s^{*+}\to e^+\nu_e)=(2.1^{+1.2}_{-0.9_{\textrm{stat.}}}\pm 0.2_{\textrm{syst.}})\times 10^{-5}$. By taking the ratio of branching fractions  $\mathcal{B}(D_s^{*+}\to e^+\nu_e)/\mathcal{B}(D_s^{+}\to \mu^+\nu_\mu)$ one obtains 
\begin{equation}
	\Gamma^{\textrm{total}}_{D_s^{*+}}=\frac{2.04\times 10^{-3}}{\mathcal{B}(D_s^{*+}\to e^+\nu_e)}\left(\frac{f_{D_s^{*+}}}{f_{D^+_s}}\right)^2\,\textrm{eV}.
\end{equation}
Using the world average $\mathcal{B}(D_s^{+}\to \mu^+\nu_\mu)$ and average value  $\frac{f_{D_s^{*+}}}{f_{D^+_s}}=1.12(1)$ given by LQCD calculations, the BESIII collaboration found the total width of $D_s^{*+}$ to be $\Gamma^{\textrm{total}}_{D_s^{*+}}=121.9^{+69.6}_{-52.2}\pm 11.8\,\textrm{eV}$~\cite{BESIII:2023zjq}, in agreement with the LQCD prediction $70(28)\,\textrm{eV}$~\cite{Donald:2013sra} within $1\sigma$. The world average value $\mathcal{B}(D^{*+}_s\to D^+_s\gamma)=0.381(29)$~\cite{Workman:2022ynf} can now be used to obtain $\Gamma(D^{*+}_s\to D^+_s\gamma)=46(30)\,\textrm{eV}$, which is much smaller than most theoretical predictions listed in Table~\ref{tab:width_VPg}, including ours. It should be noted that this estimation is based on the first measured branching fraction $\mathcal{B}(D_s^{*+}\to e^+\nu_e)=(2.1^{+1.2}_{-0.9_{\textrm{stat.}}}\pm 0.2_{\textrm{syst.}})\times 10^{-5}$ which still suffers from large uncertainties. A more precise measurement of $\mathcal{B}(D_s^{*+}\to e^+\nu_e)$, likely to be obtained in the near future, will help clarify the value of the decay width $\Gamma(D^{*+}_s\to D^+_s\gamma)$.  

\begin{figure}[ht]
	\includegraphics[scale=1.5]{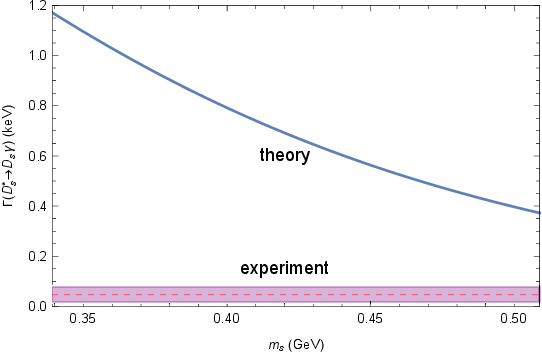}
	\caption{$\Gamma(D^*_s\to D_s\gamma)$ vs. strange-quark mass.}
	\label{fig:ms}
\end{figure}

Given that our predictions for the case of $D^{*+}$ and $D^{*0}$ fully agree with the experimental values, we wanted to check if the discrepancy in the case of $D^*_s$ is due to the value of the constituent $s$-quark mass. We varied the mass of $s$-quark by approximately $\pm 20\%$ with respect to its fitted value $m_s = 0.424$~GeV while keeping other parameters unchanged. The dependence of the width $\Gamma(D^*_s\to D_s\gamma)$ on $m_s$ is shown in Fig.~\ref{fig:ms}.  Note that our prediction is not consistent with the experimental data within the errors. More accurate experimental data are needed in the case of $D^*_s$ radiative decay. If this discrepancy still remains, we will have to refit our model parameters in the future.

Finally, we compared the contribution from the charm quark and the other one (up, down, and strange quark) into the radiative coupling constants. As mentioned above, we have that
\begin{eqnarray}
g_{D^*D\gamma}&=&e_c\,I_c+e_q\,I_q,\nonumber\\
I_{q}(m^2_{D^*},m^2_D,m_q, m_c, \omega_1,\omega_2)&=&I_{c}(m^2_{D^*},m^2_D,m_c, m_q, \omega_2,\omega_1),
\end{eqnarray}
where the expression of $I_{c(q)}$ is given in Eq.~(\ref{eq:I}). The electromagnetic aspect of the decays is mostly visible in the proportionality of the quark contribution to its electric charge. Given that $e_c=e_u=+2/3$ and $e_s=e_d=-1/3$, there is a large cancellation between the contributions from the charm and down quarks in $g_{D^{*+}D^+\gamma}$, as well as from the charm and strange quarks in $g_{D_s^{*+}D_s\gamma}$. Meanwhile, for $g_{D^{*0}D^0\gamma}$, the contributions from the charm and up quarks add up (see Table~\ref{tab:gpvg}). The strong effects in the decays are captured in the quantities $I_{c(q)}$. Table~\ref{tab:Iq} compares $I_c$ and $I_q$, $(q=u,d,s)$ by taking the ratios $I_q/I_c$. This table also includes the numerical values for the ratios $e_q I_q/e_c I_c$.

\begin{table}[ht]
	\caption{Comparison of quark contributions to radiative decay constant}
	\label{tab:Iq}
	\begin{center}
		\begin{tabular}{c|c|c|c|c}
			\hline\hline
			Channel & Quark $q$ & $I_q/I_c$ & $e_q I_q/e_c I_c$ & $g_{D^{*}D\gamma}$ (GeV$^{-1}$)\\
			\hline
			$D^{*+}\to D^+\gamma$ & $d$ & 3.74 & $-1.87$ & $-0.45(9)$  \\
			$D^{*0}\to D^0\gamma$ & $u$ & 3.74 & $+3.74$  & $+1.72(34)$\\
			$D^{*+}_s\to D^+_s\gamma$ & $s$ & 2.97 & $-1.49$ & $-0.29(6)$\\
			\hline\hline
\end{tabular}
\end{center}
\end{table}

\section{Summary}
\label{sec:sum}
Radiative decays of the vector mesons $D^{*+}$, $D^{*0}$, and $D^{*+}_s$ are studied in the framework of the Covariant Confined Quark Model in light of new experimental data from the \textit{BABAR} and BESIII collaborations. Predictions for the radiative couplings $g_{D^{*+}D^{+}\gamma}$, $g_{D^{*0}D^{0}\gamma}$, and $g_{D^{*+}_s D^{+}_s\gamma}$, as well as the decay widths are reported. A full agreement between these predictions and experimental data was found for the decays $D^{*+}\to D^{+}\gamma$ and $D^{*0}\to D^{0}\gamma$. In the case of $D^{*+}_s\to D^{+}_s\gamma$, our result for the decay width $0.55(22)\,\textrm{keV}$ 
is smaller than the recent LCSR prediction $2.36^{+1.49}_{-1.41}\,\textrm{keV}$ by more than $1\sigma$ but exceeds the first LQCD prediction $0.066(26)\,\textrm{keV}$ by more than $2\sigma$. 
Using the branching fraction of the purely leptonic decay $D_s^{*+}\to e^+\nu_e$ recently measured by the BESIII collaboration, we estimate $\Gamma(D^{*+}_s\to D^{+}_s\gamma)$ to be approximately $0.05(3)\,\textrm{keV}$. Meanwhile, theoretical predictions for this width range from $0.07\,\textrm{keV}$ to $2.4\,\textrm{keV}$, and most of them are around $0.2\,\textrm{keV}$. 
In particular, this estimated value is smaller than our prediction by more that $2\sigma$. Therefore, the decay $D^{*+}_s\to D^{+}_s\gamma$ is an interesting case that requires further theoretical and experimental studies to give a final answer.

\begin{acknowledgments}
This work is supported by Ho Chi Minh City University of Technology and Education under Grant \textnumero T2022-26. CTT deeply thanks Dr. Anh-Vu Phan-Gia and Dr. Hai-Cat Tran for their help and fruitful discussion. This paper is dedicated to the memory of our dear friend and colleague J\"urgen K\"orner.
\end{acknowledgments}

\appendix
\section{The compositeness condition} 
Let us consider the simplest case of a zero-spin bound state whose constituents are fermions. The Yukawa-type ``bare" Lagrangian describing the coupling of the boson field $\phi_0$ to the fermion field $q$ has the form
\begin{equation}
	\mathcal{L}_Y = \bar{q}(i\not\!\partial-m_q)q + \frac{1}{2}\phi_0(\Box-m_0^2)\phi_0 + g_0\phi_0(\bar{q}\Gamma q),\quad \textrm{where} \quad \Box = -\partial^\mu\partial_\mu, \quad \Gamma = I, i\gamma^5.
\end{equation}
The vacuum generating functional of the Yukawa theory is given by
\begin{equation}
	Z_Y = \int\!\mathcal{D}\phi_0\int\!\mathcal{D}\bar{q}\int\!\mathcal{D}q \,e^{i\int\! dx\mathcal{L}_Y(x)}.
\end{equation}
Integrating out quark fields and dismissing all irrelevant normalization factors, we have
\begin{align}
	Z_Y &= \int\!\mathcal{D}\phi_0 \exp  \Big\{\frac{i}{2}\int\! dx\,\phi_0(x)(\Box-m_0^2)\phi_0(x)\nn
	&- \sum_{n=1}^{\infty}\frac{i^n g_0^n}{n}\int\! dx_1\cdots\int\! dx_n\,\phi_0(x_1)\cdots\phi_0(x_n)
	\Tr[\Gamma S_q(x_1-x_2)\cdots\Gamma S_q(x_n-x_1)]\Big\},
\end{align}
where $\displaystyle S_q(x-y)=\int\frac{d^4k}{(2\pi)^4i}\frac{e^{-ik(x-y)}}{m_q-\not\!k}$ is the quark Green function (propagator).

The bi-linear terms in boson field read
\begin{align}
	L_Y^{(2)}&=\frac{1}{2}\int\! dx\,\phi_0(x)(\Box-m^2)\phi_0(x)\nn
	&-\frac{i}{2}g_0^2\int\! dx_1\int\! dx_2\,\phi_0(x_1)\phi_0(x_2)\Tr[\Gamma S_q(x_1-x_2)\Gamma S_q(x_2-x_1)]\nn
	&=\frac{1}{2}\int\! dx\,\phi_0(x)(\Box-m^2)\phi_0(x)+\frac{1}{2}g_0^2\int\! dx_1\int\! dx_2\,\phi_0(x_1)\Pi_{S=0}(x_1-x_2)\phi_0(x_2),
\end{align}
where $\displaystyle
\Pi_{S=0}(x_1-x_2) = i\left\langle T[(\bar{q}\Gamma q)_x(\bar{q}\Gamma q)_y]\right\rangle_0 = -i\,\Tr[\Gamma S_q(x_1-x_2)\Gamma S_q(x_2-x_1)]$ is the mass function of the spin-0 boson.

Expanding the Fourier transform of the mass function at the physical boson mass up to second order
\begin{equation}
	\tilde{\Pi}_{S=0}(p^2)=\int\! dx e^{-ipx}\Pi_{S=0}(x)=\tilde{\Pi}_{S=0}(m^2)+(p^2-m^2)\tilde{\Pi}^\prime_{S=0}(m^2)+\tilde{\Pi}^{\textrm{ren}}_{S=0}(p^2)
\end{equation}
we obtain
\begin{align}
	L_Y^{(2)}&=\frac{1}{2}\int\! dx\,\phi_0(x)\big[\Box-m_0^2+g_0^2\tilde{\Pi}_{S=0}(m^2)+(\Box-m^2)\tilde{\Pi}^\prime_{S=0}(m^2)\big]\phi_0(x)\nn
	&+\frac{1}{2}g_0^2\int\! dx_1\int\! dx_2\,\phi_0(x_1)\Pi^{\textrm{ren}}_{S=0}(x_1-x_2)\phi_0(x_2).
\end{align}
The wave function, boson mass, and Yukawa coupling are renormalized as follows:
\begin{eqnarray}
	\phi_r&=&Z^{-1/2}\phi_0,\qquad m^2 = m_0^2-g_0^2\tilde{\Pi}_{S=0}(m^2),\nn
	g_r &=& Z^{1/2}g_0,\qquad Z = \frac{1}{1+g_0^2\tilde{\Pi}^\prime_{S=0}(m^2)}.
\end{eqnarray}
Note that the wave function renormalization constant $Z$ is related to the renormalized coupling constant $g_r$ via 
\begin{equation}
\label{eq:Z-gr}
	Z = 1-g_r^2\tilde{\Pi}^\prime_{S=0}(m^2).
\end{equation}
The renormalized generating functional of the Yukawa theory is finally expressed as
\begin{align}
	\label{eq:ren-Yu}
	Z_Y^{\textrm{ren}}&= \int\!\mathcal{D}\phi_r \exp
	\Big\{
	\frac{i}{2}\int\! dx\,\phi_r(x)(\Box-m^2)\phi_r(x)\nn
	&+\frac{i g_r^2}{2}\int\! dx_1\int\! dx_2\,\phi_r(x_1)\Pi_{S=0}^{\textrm{ren}}(x_1-x_2)\phi_r(x_2)\nn
	&- \sum_{n=3}^{\infty}\frac{i^n g_r^n}{n}\int\! dx_1\cdots\int\! dx_n\,\phi_r(x_1)\cdots\phi_r(x_n)\Tr[\Gamma S_q(x_1-x_2)\cdots\Gamma S_q(x_n-x_1)]
	\Big\}.
\end{align}
We dismiss the linear boson term because it is absent for pseudoscalar mesons; for scalar mesons, it can be removed by a shift of the field.

Next, we move on to the Fermi theory for which the generating functional is expressed as
\begin{equation}
	Z_F = \int\! \mathcal{D}\bar{q}\int\! \mathcal{D}q e^{i\int\! dx\mathcal{L}_F(x)},
\end{equation} 
where the Fermi-type Lagrangian $\mathcal{L}_F(x)$ is given by
\begin{equation}
	\mathcal{L}_F = \bar{q}(i\not\!\partial-m_q)q + \frac{G}{2}(\bar{q}\Gamma q)^2.
\end{equation}
For the exponential of the four-fermion interaction, the Gaussian functional representation can be used as follows:
\begin{equation}
	e^{i\frac{G}{2}\langle(\bar{q}\Gamma q)^2\rangle} = N^{-1}_{+}\int\! \mathcal{D}\phi\exp 
	\big\{
	-\frac{i}{2}\frac{1}{G}\langle\phi^2\rangle + i\langle\phi\cdot (\bar{q}\Gamma q)\rangle
	\big\},
	\quad \textrm{where}\quad \langle(\cdots)\rangle = \int\! dx(\cdots).	
\end{equation}
We then have
\begin{equation}
	Z_F = N^{-1}_{+}\int\! \mathcal{D}\phi\int\!\mathcal{D}\bar{q}\int\!\mathcal{D}q\, \exp 
	\big\{
	-\frac{i}{2}\frac{1}{G}\langle\phi^2\rangle + i\langle\bar{q}[i\not\!\partial-m_q+\phi\Gamma] q\rangle
	\big\}.
\end{equation}

Integrating out quark fields and dismissing all irrelevant normalization factors, we obtain
\begin{align}
	Z_F &= \int\!\mathcal{D}\phi\exp
	\Big\{-\frac{i}{2}\frac{1}{G}\int\! dx\phi^2(x)\nn
	&- \sum_{n=1}^{\infty}\frac{i^n}{n}\int\! dx_1\cdots\int\! dx_n\phi(x_1)\cdots\phi(x_n)\Tr[\Gamma S_q(x_1-x_2)\cdots\Gamma S_q(x_n-x_1)]
	\Big\}.
\end{align}
All bi-linear terms in the boson field can be collected, and the renormalized mass function is introduced in a similar manner to that of the Yukawa case as follows:
\begin{align}
	L_F^{(2)} &= \frac{1}{2}\int\!dx\,\phi(x)
	\Big[
	-\frac{1}{G}+\tilde{\Pi}_{S=0}(m^2)+(\Box-m^2)\tilde{\Pi}^\prime_{S=0}(m^2)
	\Big]\phi(x)\nn
	&+\frac{1}{2}\int\!dx_1\int\!dx_2\,\phi(x_1)\Pi^{\textrm{ren}}_{S=0}(x_1-x_2)\phi(x_2).
\end{align}
If the condition $\displaystyle G\tilde{\Pi}_{S=0}(m^2)=1$ is required and the boson field is rescaled as $\displaystyle\phi\to\frac{\phi}{\sqrt{\tilde{\Pi}^\prime_{S=0}(m^2)}}$, one obtains the free Lagrangian of the boson field with mass $m$ and the correct residue of the Green function. The renormalized generating functional of the Fermi theory finally reads
\begin{align}
	\label{eq:ren-Fermi}
	Z_F^{\textrm{ren}} &= \int\!\mathcal{D}\phi \exp 
	\Big\{\frac{i}{2}\int\! dx\,\phi(x)(\Box-m^2)\phi(x)\nn
	&+\frac{i}{2}\frac{1}{\tilde{\Pi}^\prime_{S=0}(m^2)}\int\! dx_1\int\! dx_2\,\phi(x_1)\Pi_{S=0}^{\textrm{ren}}(x_1-x_2)\phi(x_2)\nn
	&-\sum_{n=3}^{\infty}\frac{i^n}{n}\left[1/\sqrt{\tilde{\Pi}^\prime_{S=0}(m^2)}\right]^n\int\! dx_1\cdots\int\! dx_n\,\phi(x_1)\cdots\phi(x_n)\nn
	&\times \Tr[\Gamma S_q(x_1-x_2)\cdots\Gamma S_q(x_n-x_1)]
	\Big\}.
\end{align}	 
Comparing the renormalized generating functionals of the Yukawa theory in Eq.~(\ref{eq:ren-Yu}) and the Fermi theory in Eq.~(\ref{eq:ren-Fermi}), it can be concluded that the condition for their equality is
\begin{equation}
	g_r = 1/\sqrt{\tilde{\Pi}^\prime_{S=0}(m^2)}=0.
\end{equation}
Taking into account the relation~(\ref{eq:Z-gr}), this condition is equivalent to
\begin{equation}
	Z = 1-g_r^2\tilde{\Pi}^\prime_{S=0}(m^2)=0.
\end{equation}
Therefore, the vanishing of the wave function renormalization constant in the Yukawa theory can be used as the condition for the bare (unrenormalized) field $\phi_0 = Z^{1/2}\phi_r$ to vanish for a composite boson.

\end{document}